\documentclass[titlepage,12pt]{utarticle}
\usepackage{epsf}
\usepackage{amsmath,amsfonts}
\long\def\omit#1{}

\numberwithin{equation}{section}

\begin{document}

\preprint{
UTTG--04--97\\
{\tt hep-th/9702108}\\
}

\title{M(atrix)-Theory in Various Dimensions}

\author{David Berenstein and Richard Corrado
	\thanks{Research supported in part by the Robert A.\ Welch
	Foundation and NSF Grant PHY~9511632.}}
\oneaddress{ Theory Group, Department of Physics\\ 
	University of Texas at Austin\\
  	Austin TX 78712 USA  \\ {~}\\ 
	\email{david@zippy.ph.utexas.edu}
	\email{rcorrado@zippy.ph.utexas.edu}}

\date{13 February 1997}

\Abstract{We demonstrate the precise numerical correspondence between
long range scattering of supergravitons and membranes in supergravity
in the infinite momentum frame and in M(atrix)-Theory, both in
11~dimensions and for toroidal compactifications. We also identify
wrapped membranes in terms of topological invariants of the vector
bundles associated to the field theory description of compactified
M(atrix)-Theory.  We use these results to check the realization of
T-duality in M(atrix)-Theory.}

\maketitle
\renewcommand{\baselinestretch}{1.25} \normalsize

\section{Introduction}

In the past few months, a great deal of evidence has been accumulated
in favor of the matrix model description of M-Theory proposed by
Banks, Fischler, Shenker, and Susskind~\cite{Banks:Conjecture}.  This
M(atrix)-Theory, as it has come to be known, is the large~$N$ limit of
the maximally supersymmetric quantum mechanics of $U(N)$ matrices.
According to this picture, D0-branes are partons whose bound states
represent the 11-dimensional graviton
supermultiplet~\cite{{Townsend:revisited},{Witten:variousdims}}.  

M(atrix)-Theory is formulated in the infinite momentum frame in a
compact 11-dimension, so that all modes with negative or vanishing
$p_{11}$~component\footnote{We follow~\cite{Banks:Conjecture} in using
$p_{11}$ to denote the light-cone momentum~$p_+=p_0+p_{D-1}$.} can be
integrated out, leaving only the D0-brane bound states for which 
\begin{equation}
p_{11} = \frac{N}{r},
\end{equation}
where $r$ is the radius of the 11-dimension and $N>0$ labels the number
of D0-branes in the state.  The uncompactified infinite momentum limit
is defined by taking  $N,R \rightarrow\infty$,
$N/R\rightarrow\infty$. The decoupling of anti-D0-branes is crucial in
avoiding the tachyonic divergence that appears in brane-antibrane
interactions at distances of order the string
scale~\cite{Banks:BraneAntibrane}.  The action of the theory is simply
the world-volume effective action generated by open strings stretching
between $N$~D0-branes, which can be obtained from the dimensional
reduction of ${\cal N}=1$, $D=9+1$ SYM down to
$0+1$~dimensions~\cite{{Witten:BoundStrings},{Danielsson:Dparticle},
{Kabat:Zerobrane}}. In~\cite{Banks:Conjecture}, the authors
demonstrated that the potential between two D0-branes in
M(atrix)-Theory is the same as that between 11-dimensional
supergravitons, in agreement with the worldsheet calculations
of~\cite{{Lifschytz:Compare},{Douglas:ShortDistances}}.  We
carefully reproduce this potential in 11-dimensional supergravity in
the infinite momentum frame and in M(atrix)-Theory and find exact
numerical agreement within the conventions we present.   

Toroidal compactification of M(atrix)-Theory was also explained
in~\cite{Banks:Conjecture} and elucidated in~\cite{Taylor:Compact}:
M(atrix)-Theory on $T^k$ is equivalent to $k+1$-dimensional
supersymmetric Yang-Mills theory (SYM) on the dual torus. In
particular, M(atrix)-Theory compactified on $T^3$ was considered
in~\cite{{Susskind:Tduality},{Ganor:BranesFluxes},{Sethi:RotInv}}, where it
was shown that Type~II T-duality is guaranteed by the S-duality of the
${\cal N}=4$, $D=4$~SYM that describes the system. We consider the
potential between D0-branes when we toroidally compactify each of the
spatial dimensions and again find exact agreement between
M(atrix)-Theory and supergravity, including the correct logarithmic
potential of 4-dimensional gravity in the infinite momentum frame.  

Using a construction first obtained in~\cite{deWit:QMSupermembrane},
the authors of~\cite{Banks:Conjecture} showed that M(atrix)-Theory
contains supermembranes.
In~\cite{{Aharony:MemDyn},{Lifschytz:MemInt}},  the scattering of
gravitons and membranes and membranes and anti-membranes was
considered, where it was shown that M(atrix)-Theory produces the
correct potentials of supergravity, up to an ambiguity owing to the
degeneracy of states on the membrane world-volume. We consider toroidal
compactifications with membranes wrapped on 2-cycles and find that
their contribution can be calculated using the first Chern class of
the $U(N)$~vector bundle of the SYM theory.  We again find total
agreement between M(atrix)-Theory and supergravity. We also find an
explicit realization of T-duality in the description 
of wrapped membrane states that can be identified. Furthermore, we 
expect that wrapped longitudinal 5-branes in
M(atrix)-Theory~\cite{{Berkooz:FiveBranes},{Ganor:BranesFluxes},{Banks:BranesMat}}
may be discussed in terms of the second Chern class of the bundle, an
observation that is evident in the instanton description
of~\cite{Banks:BranesMat}. 

\section{Graviton-Graviton Scattering in Eleven-Dimensional
Supergravity} 

Since M(atrix)-Theory is formulated in the infinite $p_{11}$~frame, we
want to compute the scattering amplitude in 11-dimensional
supergravity for two gravitons in the infinite momentum frame with
zero energy and~$p_{11}$ transfer. At tree level, there is no
contribution from intermediate gravitinos or antisymmetric tensors, so
the calculation is simply that of linearized gravity in $D$=11
dimensions.  A succinct expression for the tree level amplitude may be
obtained with the realization that the field theoretic amplitude is
easily extracted from the zero Regge slope limit of the Type~II
superstring four graviton amplitude~\cite{Sannan:GravityLimit}.  A
brief discussion of our conventions is given in the Appendix.
Since this amplitude contains approximately 150 terms, we will
consider the case of immediate interest, in which the gravitons have
momenta orthogonal to their polarizations, which was studied
in~\cite{{Douglas:ShortDistances},{Banks:Conjecture}}.  In this
case, the amplitude is simply 
\begin{equation} \label{eq:simptree}
A = - \frac{\kappa^2}{4} \, 
\frac{ p^{(1)}_{11} p^{(2)}_{11}}{ \left( p^{(1)}_\perp 
- p^{(4)}_\perp\right)^2 } 
\left( \frac{p^{(1)}_\perp}{p^{(1)}_{11}} 
- \frac{p^{(2)}_\perp}{p^{(2)}_{11}}\right)^4 
\end{equation}
We note that this is precisely what one would obtain from the
prescription 
\begin{equation}
A= - \kappa^2 \frac{K^{(1)} K^{(2)}}{q_\perp^2},
\end{equation}
where $K^{(i)}$ denotes the relative kinetic energy in the galilean
form appropriate to the infinite momentum frame.

We can obtain the effective graviton-graviton potential from this
scattering amplitude by a Fourier transform over the dimensions
transverse to the light cone 
\begin{equation}
V(R) = \frac{1}{2\pi r} \int \frac{ d^9q_\perp }{ (2\pi)^9 } \, A,
\end{equation}
where $1/r$ is the ``quantum'' of $p_{11}$. In $d>2$ dimensions, the
scalar Green's function is    
\begin{equation}
G_d(x) = \frac{1 }{ (d-2) \Omega_{d-1} |x|^{d-2} },
\end{equation}
where $\Omega_k$ is the surface area of the $k$-sphere (here, of
course, $d=9$). Using~(\ref{eq:kappastr}) we find the Newton law  
\begin{equation} \label{eq:newton}
\begin{split}
V(R) &= - \frac{15 }{ 16} \, \frac{ r^2 }{ T_A^3 }  \,
\frac{ p^{(1)}_{11} p^{(2)}_{11} }{ |R|^7 } \,
\left( \frac{p^{(1)}_\perp}{ p^{(1)}_{11}} 
- \frac{p^{(2)}_\perp}{ p^{(2)}_{11}}\right)^4 \\
 &= - \frac{15 }{ 16} \, \frac{ N_1 N_2 }{ T_A^3 } \, \frac{v^4}{|R|^7},
\end{split}
\end{equation}
where in the last expression we have used the definitions appropriate to
the matrix model, 
\begin{equation} \label{eq:velocity}
p_{11} = \frac{N}{r}, \qquad v= \Delta \left(
\frac{p_\perp}{p_{11}}\right).
\end{equation}

If we compactify on a $k$-torus $T^k$, we instead find that
\begin{equation} \label{eq:newtononk}
V(R) = - \frac{\pi^4}{2 {\rm Vol}(T^k)} \, \frac{N_1 N_2 v^4}{T_A^3}    
\cdot \begin{cases}
\left( (7-k) \Omega_{8-k} |R|^{7-k} \right)^{-1} & \text{for $k<7$},  \\
-\frac{1}{4\pi}\ln R^2 & \text{if $k=7$}. 
\end{cases}
\end{equation}

\section{Scattering of D0-branes in M(atrix)-Theory}

From considering the tensions of the open strings which stretch
between D0-branes one obtains the dimensionally correct
M(atrix)-Theory Lagrangian
\begin{equation} \label{eq:matrix} 
{\cal L} = \mbox{Tr} \left[ \frac{1}{2r}  ( D_t X_i )^2  
- \frac{T_A^2 }{4r} [X_i,X_j]^2 - \theta^T D_t \theta 
- \frac{T_A^2 }{4} \theta^T \gamma_i [\theta, X_i]. \right],
\end{equation}
where $D_t = \partial_t + i T_A A_0 $.  

We proceed to quantize this theory in the covariant background field
gauge with 
\begin{equation} \label{eq:backgnd}
D_{\mu B} A_\mu = 0,
\end{equation}
where $B_0$ and the $B_\alpha$ are zero for $A_0$ and the compact $X_\alpha$ and
$B_i=R_i$ for the Higgs components, corresponding to the
position vectors of the D0-branes. The
advantage of using this background gauge is that one can directly
check that $(F_{0i})^2$ vanishes at one-loop order. Therefore there is
no need for finite renormalizations, even though we are not using a
gauge that is manifestly supersymmetric. Although we can do a
supergraph calculation that respects some of the supersymmetries
(since (\ref{eq:matrix}) may be obtained from the dimensional
reduction of ${\cal N} = 4$, $D=4$~SYM to quantum mechanics), such
a method does not extend to calculations for M(atrix)-Theory
compactifications, since there is no guarantee of a superfield
formalism in more than four dimensions. In the background field method
all of the components of the gauge field, as well as the ghost fields,
couple in the same way to the background field. Therefore we can still
benefit from supersymmetry and algebraic cancellations between bosons
and fermions are still possible before doing the loop integrals. At
the one-loop level, we have results that are free from infinities.  
  
As a simplification, we will consider the interaction of 2~D0-branes,
for which the gauge group is $U(2)$. For $N$~D0-branes interacting
with  $N^\prime$~D0-branes, the group would be \\
$U(N+N^\prime)$, but we
would get the same answer multiplied by a factor $N N^\prime$ to
account for the degeneracy of the off diagonal matrix elements that
one integrates out. Additionally, we factor out the irrelevant center
of mass motion and actually consider $SU(2)$ matrices.  The one-loop
effective 4-point function obtained in this way is
\begin{equation} 
A= - 6  \int \frac{dw}{2\pi} \frac{(T_A v)^4}{(w^2+T^2_A R^2)^4}
\end{equation}
which is readily integrated to yield
\begin{equation} \label{eq:matamp}
A= - \frac{15}{16} \, \frac{1}{T_A^3} \, \frac{v^4}{R^7},
\end{equation}
which matches~(\ref{eq:newton}) exactly.  

When compactifying on $T^k$ one must consider the $k+1$~SYM on the
dual torus~\cite{Taylor:Compact}. In particular, this means that~$v$
is taken to be the relative velocity in lower dimensions, which
corresponds to the motion of the Higgs moduli in the toroidal
theory. Moreover, the integral above becomes 
\begin{equation} 
A= - 6 \sum_{n_i}  \int \frac{dw}{2\pi} \, 
\frac{(T_A v)^4}{(w^2 + T_A^2 R^2 + T^2_A (n_i e_i)^2 )^4},
\end{equation}
where the $e_i$  are the dual vectors (momentum labels) of the dual
torus, which are just the lattice vectors of the original torus.
At large $R$, we can turn the sums into integrals to obtain
\begin{equation}
A= - 6 \int \frac{dw}{2\pi} \,  \frac{dx_1\dots dx_k}{ {\rm Vol}(T^k)} \,
\frac{ (T_A)^{4-k} v^4}{( w^2 + T_A^2 R^2 +x^2 )^4}. 
\end{equation}
After integrating this in spherical coordinates, we obtain
\begin{equation} 
A= - \frac{1}{ {\rm Vol}(T^k) } \, \frac{v^4}{T_A^3} \, 
\frac{\pi^\frac{k-1}{2} \Gamma\left(\frac{7-k}{2}\right)}{4 R^{7-k}}.
\end{equation}
Noting that 
\begin{equation}
\Gamma\left(\frac{7-k}{2}\right) 
= \frac{ 4 \pi^{\frac{9-k}{2}} }{ (7-k) \Omega_{8-k} },
\end{equation}
we recover precisely~(\ref{eq:newtononk}) for $k<7$. More care must be
taken when $k=7$, since the integral obtained diverges
logarithmically.  We take 
\begin{equation} \label{eq:divergence}
\lim_{k\rightarrow 7} 
\frac{ \Gamma\left( \frac{7-k}{2} \right) }{ R^{k-7} } 
\sim \ln\Lambda - \ln R^2 
\end{equation}
and recover~(\ref{eq:newtononk}) for $k=7$.

One comment about divergences is in order. While it is true that the
covariant gauge~(\ref{eq:backgnd}) has allowed the maximal
supersymmetry of M(atrix)-Theory to work its miracles and thereby
maintain zero $\beta$-function in gauge theories in as many as~$7+1$
dimensions, we still have a divergence in~(\ref{eq:divergence}), which
we chose to regulate with the UV cutoff~$\Lambda$. This cutoff has an
obvious explanation once we recall that the SYM~theory is formulated
on the {\it dual} torus, so that UV scales are exchanged with IR
scales. Therefore $\Lambda$~is simply the cutoff associated to the IR
divergence of the supergravity logarithmic potential
in~(\ref{eq:newtononk}). Since we consider graviton-graviton
scattering at zero~$p_{11}$ transfer, this IR scale is $1/r$.   

\section{Wrapped Membranes in M(atrix)-Theory}

We would now like to consider M(atrix)-Theory on a torus~$\CT$ and
calculate the scattering of the D0-branes obtained by wrapping
supermembranes around 2-cycles in $H_2(\CT)$. The wrapped membranes
should correspond to different topological sectors of the
dimensionally reduced SYM that described the compactification. In
particular, magnetic fluxes should provide the invariants needed to
classify such configurations. A previous discussion of fluxes and
wrapped membranes is contained in~\cite{Ganor:BranesFluxes}.  One
expects to get results that agree with T-duality between the Type~II
theories. We will show that this is the case.

Let us first consider toroidal compactification to 9~dimensions,
namely when $\CT=T^2$, which is the smallest manifold upon which we
can completely wrap the membrane. Of interest to us are solutions to
the Yang-Mills equations which are time-independent and that preserve
half of the supersymmetries, {\it i.e.}\ which are BPS saturated
states. Therefore we will consider a bound state of $N$~D0-branes
which gives rise to a non-trivial background which preserves the full
rotation group. In particular, we will take all of the Higgs fields to
vanish. 

Following Atiyah and Bott~\cite{Atiyah:YMRiemann}, we will have a pure
Yang-Mills connection on the torus. Part of the invariants are the
eigenvalues of the $U(N)$ gauge field strength, which are constant
along the surface. This means that the $U(N)$~bundle splits into 
$U(N_1)\otimes U(N_2)\otimes \cdots \otimes U(N_k)$, where each of the
$U(N_i)$ represent collections of $N_i$~degenerate eigenvalues.
Moreover, each eigenvalue is quantized as given by its first Chern
class, $c_1(U(N_i))$.

To obtain configurations that only break half of the supersymmetries,
all of the eigenvalues have to be the same. Under these conditions,
all of the curvature is contained in the $U(1)$ factor of the 
$U(N) \sim SU(N)\times U(1)$ splitting~\cite{Banks:BranesMat}.  This
means that the contribution of the $SU(N)$ factor comes in the form of
Wilson lines; there is a continuum of such configurations. However,
the fact that these are connected means that, in the quantum
formalism, they correspond to zero modes of the gauge fields. A proper
quantization of these zero modes should provide momentum quantization
in the compact directions. 

We would now like to use this topological information to compute the
potential between a D0-brane and a wrapped membrane, described above
as a collection of D0-branes. The standard procedure
\cite{{Aharony:MemDyn},{Lifschytz:MemInt}} is to consider a D0-brane
at a large distance~$R$ and to integrate out the off-diagonal modes
corresponding to open strings which connect the D0-brane to the bound
state. As these off-diagonal modes are charged under the $U(1)$, 
they will generate an effective potential for the remaining light
$U(1)$~degrees of freedom. By pure dimensional analysis, this is
proportional to  
\begin{equation}
N \frac{F^4}{R^5},
\end{equation}
where $F$ denotes the curvature of the $U(1)$ bundle. A more detailed
calculation shows the full result to be equal to    
\begin{equation} \label{eq:wrappedfpot}
V(R)= B  \frac{ \left( F_{\mu\nu}\right)^4 }{R^5},
\end{equation}
where $B$ is the coefficient of $v^4/R^5$ that appears
in~(\ref{eq:newtononk}). However, we note that the field strength 
appearing in
the above equation does not have the canonical normalization of gauge 
theory that we are considering. The correct normalization is fixed by 
$F_{\mu\nu} = \left( F_{\mu\nu} \right)_{\text{can.}}/T_A$.

We note that there are contributions to the gravitational
potential~(\ref{eq:wrappedfpot}) even at zero velocity.
As $F= nF_0$, where $[F/2\pi] = c_1(U(1))$, is quantized, for $N=1$
we therefore interpret the states with different~$n$ as a membrane
wrapped $n$~times around a 2-cycle $c\in H_2(\CT)$, with a single
0-brane attached to it\footnote{We note that this expression for the
winding number agrees with that given as equation~(8.12)
of~\cite{Banks:Conjecture}.}. For a given~$N$, we will interpret this
state as a membrane wrapped $Nn$~times around the 2-cycle, with
$N$~D0-branes attached to it. As the SYM is formulated on the dual
torus~$\widetilde{\CT}$, we must consider flux quantization on the
dual 2-cycle, $\tilde{c}\in H_2(\widetilde{\CT})$. This quantization
condition is $F A_{\tilde{c}}= 2\pi n$, where $A_{\tilde{c}}$ is the area of
$\tilde{c}$.   We can therefore write the potential as  
\begin{equation} \label{eq:wrappedpot}
V(R)=  B N \frac{1}{R^5} \left( \frac{2\pi n}{T_A A_{\tilde{c}}} \right)^4.
\end{equation}

From the form of the potential~(\ref{eq:wrappedpot}), the
$n$~dependence suggests that it can be interpreted as a momentum label
for a graviton. Therefore, when $F$ can be made small, this momentum
becomes a continuum and one should be able to describe it as the
opening of a new dimension. Since the minimum value of $F$
scales with the inverse of the area of the 2-cycle and the energy is
proportional to $F^2 A_{\tilde{c}} \sim 1/ A_{\tilde{c}}$, this occurs
for only a very moderate cost in energy.  As the area, $A_c$, of the
2-cycle that gives the Kaluza-Klein description is proportional to the
reciprocal of $A_{\tilde{c}}$, this new dimension becomes important in
the effective supergravity field theory when the 
size of~$\CT$ shrinks. This is exactly what happens when wrapped
strings become light when considering T-duality of Type~IIA string
theory. We recall that these strings result from wrapping the membrane
along the 11-direction from the M-Theory point of view, so the states 
that become light are membranes wrapped around longitudinal 2-cycles
of $\CT\times S^1$. In this case, our 2-cycle doesn't correspond to a
membrane wrapped along the 11-coordinate, since this dimension is
taken to be large in M(atrix)-Theory, but if one rotates the different
2-cycles to make them match, one obtains agreement with Type~II results.  

Let us explicitly verify the above claim. By
expressing~$A_{\tilde{c}}$ in terms of the area, $A_c$, of its dual
2-cycle in~$\CT$ and 
using~(\ref{eq:twobranet}), we find that~(\ref{eq:wrappedpot}) can be
written as
\begin{equation} 
V(R)=  B N \frac{1}{R^5} \left( \frac{nT^{(2)} A_c}{1/r} \right)^4.
\end{equation}
Comparing this with the M(atrix)-Theory
identifications~(\ref{eq:velocity}), we identify
\begin{equation}
p_\perp = n N T^{(2)} A_c,
\end{equation}
which is precisely the value expected for a wrapped membrane, as we
described before. Moreover, the wrapped membrane has the
appropriate quantum numbers of a graviton, as the $U(1)$ is totally
decoupled as it corresponds to the center of mass motion. The
8~fermionic zero-modes furnish the 256-fold degeneracy, as required by
supergravity.

Brane-antibrane scattering can now be straightforwardly analyzed by
reversing the orientation of~$F$ on a second block of branes. As the 
non-diagonal blocks are charged with opposite signs under both
$U(1)$s, the net result will be proportional to 
\begin{equation}
N N^\prime \frac{(F-F^\prime)^4}{R^5}.
\end{equation}
Note that when the field strengths match, we have the
case of parallel branes and the potential indeed vanishes, as required
by the BPS condition. As the simplest example of a configuration for
which a potential exists, we take $n=-n^\prime=1$ and obtain a result
that is 16~times that for the interaction between a D0-brane and the
wrapped brane. This reproduces the correct potential between wrapped
branes. The same potential was analyzed
in~\cite{{Aharony:MemDyn},{Lifschytz:MemInt}} for infinite membranes,
where an ambiguity as to the correct counting of the degeneracies of
membrane states was found.  If one treats the 
constant~$F$ exactly in the Hamiltonian and disregards the effects of the 
boundary of the torus, one obtains a splitting between bosons and
fermions in the energy levels~\cite{Aharony:MemDyn}. 
To obtain the correct potential, one has to take into account the
degeneracy of these Landau levels\footnote{We thank M.\ Berkooz for
emphasizing this point to us.}, leading to an extra factor of~$n$. 
This reproduces the $F^4$~interaction we had before, and it also shows
the appearance of a tachyon in the spectrum for small separations, in
agreement with the results previously obtained
in~\cite{{Aharony:MemDyn},{Lifschytz:MemInt}}.

On compactifying further dimensions, the wrapped membrane states that
we have described remain BPS saturated solutions of the appropriate
SYM theory.  The same arguments we present above can be carried out
and the only modifications are in the structure of the second homology
groups and in the value of the gravitational constant and $R$~dependence.

Now that we have seen that we can correctly produce all of the BPS
saturated states, it is interesting to investigate non-BPS saturated
solutions.  Recall that, in general, we have the splitting
$U(N_1)\otimes U(N_2)\otimes \cdots \otimes U(N_k)$, so these states
can be interpreted as bound states of wrapped membranes.  Since a
tachyon develops in the theory at short-distances, these states are
unstable. However, the fact that they are classical solutions to the
Yang-Mills equations means that they should provide the intermediate
states in studies of brane-antibrane annihilation and in brane
scattering with transfer of RR charge. We hope to return to these
considerations in a future work.

If we compactify two more dimensions, we take~$\CT=T^4$ and have
$4+1$-dimensional SYM on the dual torus~$\widetilde{\CT}$. Associated to
this gauge theory, we now have a new quantum number, namely the
instanton number for the $SU(N)$~factor, which is the second Chern
class, $c_2(SU(N))$. A construction analogous to that above introduces
an object that is independent of the D0-branes and membranes we had
before. In the IIA~picture, this must correspond to D4-branes wrapped
around a 4-cycle, a description already given
by~\cite{Banks:BranesMat}. Such a topological analysis of D4-brane
bound states appears in~\cite{Guralnik:Torons}. Again, with  
$c_2(SU(N)) \in H^4(\CT,\BZ)$, this result generalizes
straightforwardly to higher compactifications.  

Moreover, one may
conjecture that the third Chern class should give some notion of
wrapped six-branes in Type~IIA compactified to four dimensions.

\section{Conclusions}

In this letter, we have shown that the gravitational interactions of
M(atrix)-Theory correspond exactly to those of supergravity, in flat
space and for all toroidal compactifications down to four flat
dimensions.  We have also given a numerically precise description of
membranes which are wrapped on 2-cycles of tori,  exactly
reproducing the supergravity interactions of membranes. The behavior
of these wrapped membranes is consistent with an explicit realization of 
T-duality in M(atrix)-Theory.

It appears that the physics of wrapped membranes in M(atrix)-Theory is
completely contained in a sum over the different topological sectors of
the matrix SYM describing the system.  Considering that the
quantization of membranes is itself a non-trivial matter, it is rather
remarkable that the problem is relatively tame here. The topological
properties of M(atrix)-Theory that we have uncovered should prove to
be of some importance in discussing compactifications on non-trivial
manifolds~\cite{{Douglas:EnhancedMat},{Motl:Boundaries},{Kachru:GaugeBosons},{Kim:Orbifold}}.
In particular, it would be very interesting if one could generalize
our results to non-toroidal compactifications.  We are currently
working on these issues.

\appendix
\stepcounter{section}
\section*{Conventions for M-Theory and 11-Dimensional Gravity}

We would like to establish the precise conventions that produce exact
agreement between M-Theory and M(atrix)-Theory.  We take the
coefficient of the  Einstein action to be
\begin{equation}
S_{\rm grav} = \frac{1}{ 2\kappa^2} \int d^{11}x \sqrt{g} R
\end{equation}
and in linearization use the graviton normalization 
($g_{\mu\nu} = \eta_{\mu\nu} + \kappa h_{\mu\nu}$) 
\begin{equation}
h_{\mu\nu} =  \frac{1}{\sqrt{2k_{11}}}e_{\mu\nu}(k) e^{ik\cdot x},
\end{equation}
where $|e_{\mu\nu}|^2=1$. To calculate $\kappa^2$ we use the Schwarz
formula for the M2 and M5-brane tensions~\cite{Schwarz:Power},
\begin{equation} \label{eq:schwarz} 
T^{(5)} = \frac{ \left( T^{(2)} \right)^2 }{ 2\pi },
\end{equation}
and the quantization condition~\cite{Duff:BlackDiverse}
\begin{equation} \label{eq:diracqc}
2\kappa^2 T^{(2)} T^{(5)} \in 2\pi \BZ,
\end{equation}
so that we can express
\begin{equation} \label{eq:kappasq}
\kappa^2 = \frac{ 2\pi^2 }{  \left( T^{(2)} \right)^3 }.
\end{equation}
Since the IIA string is obtained by wrapping the M2-brane around the
11th dimension, 
\begin{equation} \label{eq:twobranet}
2\pi r T^{(2)} = T_A = \frac{M_s^2 }{ 4\pi},
\end{equation}
so that we can alternatively express $\kappa^2$ in the string units
of~(\ref{eq:matrix}) 
\begin{equation} \label{eq:kappastr}
\kappa^2 = \frac{ (2\pi)^5 }{ 2 } \, 
\left( \frac{r }{ T_A} \right)^3.
\end{equation}

\section*{Acknowledgements}
We would like to thank Willy Fischler for suggesting and guiding us
along this line of research, as well as for many stimulating
discussions about M(atrix)-Theory in general. We also thank Moshe
Rozali, Vadim Kaplunovsky, Micha Berkooz, and especially Jacques
Distler for useful discussions.  


\bibliography{various}

\begingroup\raggedright\begin{thebibliography}{10}

\bibitem{Banks:Conjecture}
T.~Banks, W.~Fischler, S.~H. Shenker, and L.~Susskind, ``{M} {T}heory as a
  {M}atrix {M}odel: {A} {C}onjecture,''
  \href{http://xxx.lanl.gov/abs/hep-th/9610043}{{\tt hep-th/9610043}}.

\bibitem{Townsend:revisited}
P.~K. Townsend, ``The {E}leven-{D}imensional {S}upermembrane {R}evisited,''
  {\em Phys. Lett.} {\bf B350} (1995) 184--187,
  \href{http://xxx.lanl.gov/abs/hep-th/9501068}{{\tt hep-th/9501068}}.

\bibitem{Witten:variousdims}
E.~Witten, ``String {T}heory {D}ynamics in {V}arious {D}imensions,'' {\em Nucl.
  Phys.} {\bf B443} (1995) 85--126,
  \href{http://xxx.lanl.gov/abs/hep-th/9503124}{{\tt hep-th/9503124}}.

\bibitem{Banks:BraneAntibrane}
T.~Banks and L.~Susskind, ``Brane-{A}ntibrane {F}orces,''
  \href{http://xxx.lanl.gov/abs/hep-th/9511194}{{\tt hep-th/9511194}}.

\bibitem{Witten:BoundStrings}
E.~Witten, ``Bound {S}tates of {S}trings and $p$-{B}ranes,'' {\em Nucl. Phys.}
  {\bf B460} (1996) 335--350,
  \href{http://xxx.lanl.gov/abs/hep-th/9510135}{{\tt hep-th/9510135}}.

\bibitem{Danielsson:Dparticle}
U.~H. Danielsson, G.~Ferretti, and B.~Sundborg, ``D-particle {D}ynamics and
  {B}ound {S}tates,'' {\em Int. J. Mod. Phys.} {\bf A11} (1996) 5463--5478,
  \href{http://xxx.lanl.gov/abs/hep-th/9603081}{{\tt hep-th/9603081}}.

\bibitem{Kabat:Zerobrane}
D.~Kabat and P.~Pouliot, ``A {C}omment on {Z}erobrane {Q}uantum {M}echanics,''
  {\em Phys. Rev. Lett.} {\bf 77} (1996) 1004--1007,
  \href{http://xxx.lanl.gov/abs/hep-th/9603127}{{\tt hep-th/9603127}}.

\bibitem{Lifschytz:Compare}
G.~Lifschytz, ``Comparing {D}-branes to {B}lack-branes,'' {\em Phys. Lett.}
  {\bf B388} (1996) 720--726,
  \href{http://xxx.lanl.gov/abs/hep-th/9604156}{{\tt hep-th/9604156}}.

\bibitem{Douglas:ShortDistances}
M.~R. Douglas, D.~Kabat, P.~Pouliot, and S.~H. Shenker, ``D-branes and {S}hort
  {D}istances in {S}tring {T}heory,''
  \href{http://xxx.lanl.gov/abs/hep-th/9608024}{{\tt hep-th/9608024}}.

\bibitem{Taylor:Compact}
W.~Taylor, ``D-brane {F}ield {T}heory on {C}ompact {S}paces,''
  \href{http://xxx.lanl.gov/abs/hep-th/9611042}{{\tt hep-th/9611042}}.

\bibitem{Susskind:Tduality}
L.~Susskind, ``T {D}uality in {M}(atrix) {T}heory and {S} {D}uality in {F}ield
  {T}heory,'' \href{http://xxx.lanl.gov/abs/hep-th/9611164}{{\tt
  hep-th/9611164}}.

\bibitem{Ganor:BranesFluxes}
O.~J. Ganor, S.~Ramgoolam, and W.~Taylor, ``{B}ranes, {F}luxes and {D}uality in
  {M}(atrix)-{T}heory,'' \href{http://xxx.lanl.gov/abs/hep-th/9611202}{{\tt
  hep-th/9611202}}.

\bibitem{Sethi:RotInv}
S.~Sethi and L.~Susskind, ``Rotational {I}nvariance in the {M}(atrix)
  {F}ormulation of {T}ype {IIB} {T}heory,''
  \href{http://xxx.lanl.gov/abs/hep-th/9702101}{{\tt hep-th/9702101}}.

\bibitem{deWit:QMSupermembrane}
B.~de~Wit, J.~Hoppe, and H.~Nicolai, ``On the {Q}uantum {M}echanics of
  {S}upermembranes,'' {\em Nucl. Phys.} {\bf B305} (1988) 545.

\bibitem{Aharony:MemDyn}
O.~Aharony and M.~Berkooz, ``Membrane {D}ynamics in {M}(atrix) {T}heory,''
  \href{http://xxx.lanl.gov/abs/hep-th/9611215}{{\tt hep-th/9611215}}.

\bibitem{Lifschytz:MemInt}
G.~Lifschytz and S.~D. Mathur, ``Supersymmetry and {M}embrane {I}nteractions in
  {M}(atrix) {T}heory,'' \href{http://xxx.lanl.gov/abs/hep-th/9612087}{{\tt
  hep-th/9612087}}.

\bibitem{Berkooz:FiveBranes}
M.~Berkooz and M.~R. Douglas, ``Five-branes in {M}(atrix) {T}heory,''
  \href{http://xxx.lanl.gov/abs/hep-th/9610236}{{\tt hep-th/9610236}}.

\bibitem{Banks:BranesMat}
T.~Banks, N.~Seiberg, and S.~Shenker, ``Branes from {M}atrices,''
  \href{http://xxx.lanl.gov/abs/hep-th/9612157}{{\tt hep-th/9612157}}.

\bibitem{Sannan:GravityLimit}
S.~Sannan, ``{G}ravity as the {L}imit of the {T}ype-{II} {S}uperstring
  {T}heory,'' {\em Phys. Rev.} {\bf D34} (1986) 1749.

\bibitem{Atiyah:YMRiemann}
M.~F. Atiyah and R.~Bott, ``The {Y}ang-{M}ills {E}quations {O}ver {R}iemann
  {S}urfaces,'' {\em Phil. Trans. R. Soc. Lond. A} {\bf 308} (1982) 523--615.

\bibitem{Guralnik:Torons}
Z.~Guralnik and S.~Ramgoolam, ``Torons and {D}-brane {B}ound {S}tates,''
  \href{http://xxx.lanl.gov/abs/hep-th/9702099}{{\tt hep-th/9702099}}.

\bibitem{Douglas:EnhancedMat}
M.~R. Douglas, ``Enhanced {G}auge {S}ymmetry in {M}(atrix) {T}heory,''
  \href{http://xxx.lanl.gov/abs/hep-th/9612126}{{\tt hep-th/9612126}}.

\bibitem{Motl:Boundaries}
L.~Motl, ``Quaternions and {M}(atrix) {T}heory in {S}paces with {B}oundaries,''
  \href{http://xxx.lanl.gov/abs/hep-th/9612198}{{\tt hep-th/9612198}}.

\bibitem{Kachru:GaugeBosons}
S.~Kachru and E.~Silverstein, ``On {G}auge {B}osons in the {M}atrix {M}odel
  {A}pproach to {M} {T}heory,''
  \href{http://xxx.lanl.gov/abs/hep-th/9612162}{{\tt hep-th/9612162}}.

\bibitem{Kim:Orbifold}
N.~Kim and S.-J. Rey, ``M(atrix) {T}heory on an {O}rbifold and {T}wisted
  {M}embrane,'' \href{http://xxx.lanl.gov/abs/hep-th/9701139}{{\tt
  hep-th/9701139}}.

\bibitem{Schwarz:Power}
J.~H. Schwarz, ``The {P}ower of {M} {T}heory,'' {\em Phys. Lett.} {\bf B367}
  (1996) 97--103, \href{http://xxx.lanl.gov/abs/hep-th/9510086}{{\tt
  hep-th/9510086}}.

\bibitem{Duff:BlackDiverse}
M.~J. Duff and J.~X. Lu, ``Black and {S}uper p-branes in {D}iverse
  {D}imensions,'' {\em Nucl. Phys.} {\bf B416} (1994) 301--334,
  \href{http://xxx.lanl.gov/abs/hep-th/9306052}{{\tt hep-th/9306052}}.

\end{thebibliography}\endgroup
\bibliographystyle{utphys}

\end{document}